\documentstyle[amssymb,prl,aps,twocolumn]{revtex}
\begin{document}
\draft
\title{Local distinguishability of quantum states and the distillation of
entanglement}
\author{Ping-Xing Chen$^{1,2\thanks{%
E-mail: pxchen@nudt.edu.cn}}$and Cheng-Zu Li$^1$}
\address{$^1$ Department of Applied Physics, National University of Defense\\
Technology, Changsha, 410073, P. R. China. \\
$^2$Key Laboratory of Quantum Information, University of Science and\\
Technology of China, Chinese Academy of Sciences, Hefei 230026, P. R. China}
\maketitle

\begin{abstract}
This paper tries to probe the relation between the local distinguishability
of orthogonal quantum states and the distillation of entanglement. An new
interpretation for the distillation of entanglement and the
distinguishability of orthogonal quantum states in terms of information is
given, respectively. By constraining our discussion on a special protocol we
give a necessary and sufficient condition for the local distinguishability
of the orthogonal pure states, and gain the maximal yield of the distillable
entanglement. It is shown that the information entropy, the locally
distinguishability of quantum states and the distillation of entanglement
are closely related.
\end{abstract}


One of interesting topics in quantum mechanics is how to distinguish a set
of quantum states by local operations and classical communication (LOCC).
Alice and Bob share a quantum system, in one of a known set of possible
orthogonal states $\left| \Psi _1\right\rangle ,\left| \Psi _2\right\rangle
,...,\left| \Psi _i\right\rangle ,...,\left| \Psi _n\right\rangle $. They do
not, however, know particular state they actually possesses.. To distinguish
these possible states they should perform some sequence of LOCC. If these
states are not orthogonal to each other, they cannot be distinguished
deterministically. Further more, if these states are orthogonal to each
other, when only a single copy is provided, they still cannot be
distinguished by LOCC except for some special cases \cite{1,2,3}. Some
interesting works on locally distinguishability of quantum states have been
presented \cite{1,2,3,4,5}. For example, any three of the four Bell states

\begin{eqnarray}
\left| \Phi ^{\pm }\right\rangle &=&\frac 1{\sqrt{2}}(\left| 00\right\rangle
\pm \left| 11\right\rangle ) \\
\left| \Psi ^{\pm }\right\rangle &=&\frac 1{\sqrt{2}}(\left| 01\right\rangle
\pm \left| 10\right\rangle )  \nonumber
\end{eqnarray}
cannot be distinguished by LOCC if only a single copy is provided \cite{2}.

Another interesting topic in quantum mechanics is the distillation of
entanglement. Maximally entangled states may have many applications in
quantum information, such as error correcting code\cite{6}, dense coding\cite
{7} and teleportation\cite{8}, etc. In the laboratory, however, a maximally
entangled state always becomes a mixed state easily due to the interaction
with environment. This results in poor applications. The idea of the
distillation of entanglement is to get some maximally entangled states by
LOCC from many \cite{chen} or infinite copies of a mixed state. A few of
protocols for the distillation of entanglement were given\cite{6,9,66}. But
what is the most efficient distillation protocol and how to calculate the
distillable entanglement (the maximal value of entanglement gained from per
mixed state), $E_D,$ are still open questions.

All protocols for the distillation of entanglement have a common feature:
the distillable entanglement of a mixed state is not bigger than the
entanglement of formation of the mixed state owing to the loss of information%
\cite{10,12}. In essence, indistinguishability of a set of orthogonal states
is also owing to the loss of information. The transformation of information
plays an important role in both the distillation of entanglement and the
distinguishability of orthogonal quantum states. In this sense, the
distillation of entanglement and the distinguishability of orthogonal
quantum states should have some links. In this paper, we try to probe this
question. Closely related to the present paper is the work of Vedral and
Plenio et al \cite{v,m} who mentioned the link between the global
distinguishability of quantum states and the distillation of entanglement,
and the work in Refs.\cite{10,12} which discussed the relations of the
classical information and the entanglement. But these paper did not look at
the notion of local distinguishability.

In the asymptotic cases a protocol for the distillation of entanglement is
to get pure entangled states by LOCC from $n(n\longrightarrow \infty )$
copies of a mixed state $\sigma ,$

\begin{equation}
\sigma =\sum_{i=1}^m\lambda _i\left| \Phi _i\right\rangle \left\langle \Phi
_i\right| ,\quad \sum_{i=1}^m\lambda _i=1,  \label{a}
\end{equation}
where $\left| \Phi _i\right\rangle s$ are the eigenstates of $\sigma $ with
nonzero eigenvalues $\lambda _is.$ As shown in the paper by Bennett et al 
\cite{9,b3} that $\sigma ^{\otimes n}$ has $2^{nS(\sigma )}$ ``likely''
strings of orthogonal pure states. Because what is the most efficient
distillation protocol and how to distinguish a general set of orthogonal
states are still open questions, we first constrain our discussion on a
special protocol (which we define as {\it one by one measurement} in the
following). Under the special protocol we give the necessary and sufficient
condition for the distinguishability of the $2^{nS(\sigma )}$ ``likely''
strings of orthogonal pure states, and gain the maximal yield of the
distillable entanglement. It is shown that the information entropy, the
locally distinguishability of orthogonal quantum states and the distillation
of entanglement have close links. Finally, we consider the generalization of
the links to general protocols briefly.

In this paper we will apply the following fact in many cases.

{\bf Fact}: Suppose Alice and Bob share a pair particles whose state is $%
\sigma .$ The $n$ copies of $\sigma ,\sigma ^{\otimes n},$ is a mixture of $%
m^n$ pure states-strings, but there are only $\prod_{i=1}^mC_{n-n%
\sum_{j=0}^{i-1}\lambda _j}^{n\lambda _i}$ ``likely'' strings of orthogonal
pure states \cite{9,b3}, such as, one of the ``likely'' strings

\begin{equation}
\stackrel{\lambda _1n}{\overbrace{\left| \Phi _1\right\rangle \cdots \left|
\Phi _1\right\rangle }}\stackrel{\lambda _2n}{\overbrace{\left| \Phi
_2\right\rangle \cdots \left| \Phi _2\right\rangle }}\cdots \stackrel{%
\lambda _mn}{\overbrace{\left| \Phi _m\right\rangle \cdots \left| \Phi
_m\right\rangle }},
\end{equation}
where we note $\lambda _0=0.$ In each of ``likely'' strings there are $%
\lambda _in$ pairs whose states are $\left| \Phi _i\right\rangle .$ The
probability that each ``likely'' string occurs is $\prod_{i=1}^m\lambda
_i^{n\lambda _i}.$ It can be proved that as $n\rightarrow \infty $ we have
limits$,$

\begin{equation}
\prod_{i=1}^mC_{n-n\sum_{j=0}^{i-1}\lambda _j}^{n\lambda _i}=2^{nS(\sigma )}
\label{4}
\end{equation}
and

\begin{equation}
\prod_{i=1}^m\lambda _i^{n\lambda
_i}\prod_{i=1}^mC_{n-n\sum_{j=0}^{i-1}\lambda _j}^{n\lambda _i}=1,
\end{equation}
where $S(\sigma )$ is the information entropy of $\sigma $

\begin{equation}
S(\sigma )=-\sum_{i=1}\lambda _i\ln \lambda _i  \label{5}
\end{equation}
It is to say that the sum of probability of all ``likely'' strings tends to
1. So we only need consider the ``likely'' strings as $n\rightarrow \infty .$

Any protocol for distillation of entanglement from $n$ copies of a mixed
state $\sigma ,\sigma ^{\otimes n},$ can be conceived as successive rounds
of measurements and communication by Alice and Bob. After N rounds of
measurements and communication, there are many possible outcomes which
correspond to many measurement operators $\{A_i\otimes B_i\}$ acting on the
Alice and Bob's Hilbert space. Each of these operators is a product of the
positive operators and unitary maps corresponding to Alice's and Bob's
measurement and rotations, and represents the effect of the N measurements
and communication. If the outcome $i$ occurs, the given state $\sigma
^{\otimes n}$ becomes:

\begin{equation}
A_i^{+}\otimes B_i^{+}\sigma ^{\otimes n}A_i\otimes B_i
\end{equation}
If the state $\sigma $ is distillable, there must be at least an element $%
A_i\otimes B_i$ such that as $n\rightarrow \infty $

\begin{equation}
A_i^{+}\otimes B_i^{+}\sigma ^{\otimes n}A_i\otimes B_i\rightarrow \left|
\Psi _i\right\rangle \left\langle \Psi _i\right| ,  \label{ed}
\end{equation}
where $\left| \Psi _i\right\rangle $ is a pure entangled state in subspace $%
V_i\otimes V_i.$ The distillable entanglement of $\sigma $ is the maximum
numbers, $E_D(\sigma )$ such that there exists a set of operations as $%
n\rightarrow \infty ,$ we have limits \cite{b8}

\begin{equation}
E_D(\sigma )=\lim_{n\rightarrow \infty }\frac 1n\sum_ip_iE(\left| \Psi
_i\right\rangle ),  \label{1}
\end{equation}
where $E(\left| \Psi _i\right\rangle )$ is the entanglement of pure state $%
\left| \Psi _i\right\rangle ,$ $p_i$ is the probability Alice and Bob carry
out operation $A_i\otimes B_i.$ One of the effects of $A_i\otimes B_i$ is to
project out the subspace on which the projection of $\sigma ^{\otimes n}$ is
a pure entangled state. We define the subspace as{\normalsize \ }{\it %
distillable subspace }(DSS) \cite{chen}. In general, there are many DSS in
the Hilbert space of $n$ pairs as $n\rightarrow \infty .$

{\bf Definition: one by one measurement}{\it , the measurement is one by one
measurement if Alice and Bob measure some pairs and only measure a pair
particles at one time.}

Now we consider a protocol. The aim of the protocol is to distinguish
deterministically the $2^{nS(\sigma )}$ ``likely'' strings by Alice's and
Bob's {\it one by one measurements}. To distinguish deterministically the $%
2^{nS(\sigma )}$ ``likely'' strings, Alice and Bob should exclude the
possibility of the other ``likely'' strings and keep only a string. In terms
of information the procedure of distinguishing these ``likely'' strings is
to clear up the uncertainty of the $n$ pairs particles, or get the $%
nS(\sigma )$ bits information by rounds of local unitary operations (LUO)
and measurements each of which will destroy some entanglement of each
string. By measuring some pairs Alice and Bob can divide the $2^{nS(\sigma
)} $ strings into many strings-groups and then get some information of the $n
$ pairs. Each of the strings-groups can be distinguished from others, since
each of the groups can be ``indicated'' by the product vectors of the
measured pairs. For example, if $\sigma $ is a Bell-diagonal states Alice
and Bob can measure a pair with product bases $\left| 00\right\rangle
,\left| 01\right\rangle ,\left| 10\right\rangle ,\left| 11\right\rangle .$
Alice and Bob may get bases $\left| 00\right\rangle $ and $\left|
11\right\rangle $ with same probability $(\lambda _1+\lambda _2)/2.$ The
bases both $\left| 00\right\rangle $ and $\left| 11\right\rangle $ indicate
the strings in which the state of the measured pair is $\left| \Phi
\right\rangle ;$ Or Alice and Bob may get bases $\left| 01\right\rangle $
and $\left| 10\right\rangle $ with same probability $(\lambda _3+\lambda
_4)/2,$ and both $\left| 01\right\rangle $ and $\left| 10\right\rangle $
indicate the strings in which the state of the measured pair is $\left| \Psi
\right\rangle .$ After measuring some pairs each of the $2^{nS(\sigma )}$
strings may be indicated by the product vectors of the measured pairs and
can be distinguished. If the $2^{nS(\sigma )}$ can be distinguished we say
that Alice and Bob get the $nS(\sigma )$ bits information.

Obviously, Alice and Bob should measurement n pairs to get the $nS(\sigma )$
bits information by measurement directly on n copies without the help of any
LUO. Fortunately, it is possible that Alice and Bob get the $nS(\sigma )$
bits information by measuring less than n pairs with the help of a set of
LUO acting on the all copies $\sigma ^{\otimes n}$ and classical
communication. So, in essence, the above protocol is to distinguish the $%
2^{nS(\sigma )}$ strings by distinguishing the states of each measured pair
with the help of LUO and classical communications.

Suppose Alice and Bob need at least to measure ($n-m)$ pairs particles to
get $nS(\sigma )$ bits informations. After measuring $n-m$ pairs with the
help of a set of local unitary transformations, Alice and Bob can
distinguish the $2^{nS(\sigma )}$ strings and the entanglement of unmeasured
pairs in each string is kept. So they get a yield of entanglement,

\begin{equation}
E_D^{\prime }=\frac{mE(\sigma )}n,  \label{ss}
\end{equation}
where $mE(\sigma )$ is the entanglement of kept pairs in a string.

Operations to distinguish the states of a pair particles can be achieved by
measuring a pair with a set of product vectors \cite{3}. Suppose that $P_j$
is the probability of Alice and Bob getting the j'th product vector, and $%
P_j^{^{\prime }}$ is the sum of the probability such that the j'th product
vector indicates $P_j^{^{\prime }}2^{nS(\sigma )}$ ``likely'' strings. If
Alice and Bob get the j'th product vector, they keep $P_j^{^{\prime
}}2^{nS(\sigma )}$ strings and discard the others. In terms of information,
they get $-\ln P_j^{\prime }$ bit information. We define $-\ln P_j^{\prime }$
as {\it distinguishable information} (DI) which reflects on the contribution
to distinguish the ``likely'' strings when Alice and Bob get the j'th
product vector. If the j'th vector indicates a few of the ``likely'' strings
(or indicates a strings-group), not all the ``likely'' strings, it presents
nonzero DI, $-\ln P_j^{^{\prime }},$ and makes a contribution to distinguish
these ``likely'' strings. If a vector indicates all the ``likely'' strings,
it presents no DI, which corresponds to inability to distinguish the
``likely'' strings. Only when the DI gained by measuring some pairs is equal
to the information entropy of the $2^{nS(\rho )}$ strings, $nS(\rho ),$ can
these ``likely'' strings be distinguished. Because all strings have same
structure, by the symmetry when $n\rightarrow \infty ,$ Alice and Bob can
get the same DI from each measured pair.

Suppose that Alice and Bob have measured M pairs, we consider a kind of
outputs in which $M_j$ pairs collapse the j'th basis. The probability that
Alice and Bob get one of this kind of outputs is

\[
\prod_jP_j^{M_j}, 
\]
where $M=\sum_jM_j.$ The number of these outputs is 
\[
\prod_jC\ _{M-\sum_{i=0}^{j-1}M_i}^{M_j}, 
\]
where $M_0=0.$ Each of these outputs results in $-\ln \prod_jP_j^{\prime
M_j} $ bits informations. From the similar statement as the Fact it follows
that the ``likely'' outputs are those in which $MP_j$ pairs collapse the
j'th vector, and the probability of all the ``likely'' outputs tends to 1 as 
$n\rightarrow \infty .$ A ``likely'' output results in $-\ln
\prod_jP_j^{\prime MP_j}$ bits DI. When DI is equal to the information
entropy of n pairs, i.e.,

\begin{equation}
-\ln \prod_jP_j^{\prime MP_j}=-M\sum_jP_j\ln P_j^{^{\prime }}=nS(\sigma ),
\label{6}
\end{equation}
Alice and Bob can distinguish the $2^{nS(\sigma )}$ strings, and get a yield

\begin{equation}
E_D^{\prime \prime }(\sigma )=\frac 1n(n-\frac{nS(\sigma )}{I_d(\sigma )}%
)E(\sigma )=(1-\frac{S(\sigma )}{I_d(\sigma )})E(\sigma ),
\end{equation}
where $I_d(\sigma )=-\sum_jP_j\ln P_j^{^{\prime }},$ is a average DI by
measuring a pair. If the maximal average DI is $I_{d\max }(\sigma ),$ the
yield is

\begin{equation}
E_D^{\prime \prime \prime }(\sigma )=(1-\frac{S(\sigma )}{I_{d\max }(\sigma )%
})E(\sigma ).  \label{dd}
\end{equation}

The discussion above means that under one by one measurement protocol Alice
and Bob should measure $\frac{nS(\sigma )}{I_{d\max }(\sigma )}$ pairs at
least to get a yield in Eq. (\ref{dd}). By measuring $\frac{nS(\sigma )}{%
I_{d\max }(\sigma )}$ pairs Alice and Bob can get the all ``likely''
outputs, each of which results in a yield in Eq. (\ref{dd}). So the yield of
entanglement in equation (\ref{dd}) is the maximal yield under one by one
measurement protocol, and is a lower bound of the distillable entanglement.
On the other hand, the above discussion shows that under the one by one
measurement protocol Alice and Bob should measure $\frac{nS(\sigma )}{%
I_{d\max }(\sigma )}$ pairs at least to distinguish deterministically the $%
2^{nS(\rho )}$ ``likely'' strings, the $2^{nS(\rho )}$ ``likely'' strings
are distinguishable if and only if the yield $E_D^{\prime \prime \prime
}(\sigma )$ in Eq. (\ref{dd}) is bigger than or equal to zero, i.e., 
\begin{equation}
I_{d\max }(\sigma )\geqslant S(\sigma ).
\end{equation}
This shows a close link between the locally distinguishability of orthogonal
quantum states and the distillation of entanglement. This link is fit to all
multi-partite states.

It is well known that there are a few of upper bound of the distillable
entanglement, such as the relative entropy of entanglement \cite{v}. Here we
present a lower bound of the distillable entanglement as Eq. (\ref{dd}). If
the mixed state $\sigma $ is a Bell-diagonal state $\rho ,$ the maximal DI
is not less than 1 as shown in the Ref.\cite{6,9}, i.e., 
\begin{equation}
I_{d\max }(\rho )\geqslant 1.
\end{equation}
Given that $E(\rho )$ in Eq. (\ref{dd}) is equal to 1, we can get a lower
bound of the distillable entanglement of a Bell-diagonal state $\rho ,$

\[
E_D(\rho )\geqslant 1-S(\rho ) 
\]
Suppose that the mixed state $\sigma $ is a multiple copies of four Bell
states \cite{21}, i.e.,

\[
\sigma =\rho ^{(n)}=\frac 14\sum_{i=1}^4(\left| \Phi _i\right\rangle
\left\langle \Phi _i\right| )^{\otimes n},
\]
where $\left| \Phi _{1,2}\right\rangle =\left| \Phi ^{\pm }\right\rangle
;\left| \Phi _{3,4}\right\rangle =\left| \Psi ^{\pm }\right\rangle .$
Because a copy of four Bell states provides at least 1 bit DI, the following
inequality should hold 
\begin{equation}
I_{d\max }(\rho ^{(n)})\geqslant n.
\end{equation}
Given that $E(\rho ^{(n)})$ in Eq. (\ref{dd}) is equal to $n$, and $S(\rho
^{(n)})=2$, we can get a lower bound of the distillable entanglement of a
Bell-diagonal state $\rho ^{(n)},$

\begin{equation}
E_D(\rho ^{(n)})\geqslant n-2.
\end{equation}
On the other hand, the relative entropy of entanglement of $\rho ^{(n)}$ is
equal to $n-2,$ as shown in the Ref. \cite{21}, so we follow that $E_D(\rho
^{(n)})=n-2.$

The example above shows that the Eq. (\ref{dd}) may be useful to calculate
the distillable entanglement or the lower bound of the distillable
entanglement. But the novelty of the Eq. (\ref{dd}) is to show the close
relation among the distillation of entanglement, the local
distinguishability of orthogonal quantum states and the information entropy.

Now we would like to discuss the more general protocol briefly. To
distinguish the $2^{nS(\rho )}$ ``likely'' strings, Alice and Bob should do
rounds of measurements and classical communication. The effect of these
measurements and classical communication can be represented as a set of
operators $\{A_i\otimes B_i\}$. If the output is $i$ Alice and Bob know they
have got the $i^{\prime }th$ string with certainty, i.e.,

\begin{eqnarray}
A_i\otimes B_i\left| string_i\right\rangle &=&\left| string_i^{\prime
}\right\rangle ; \\
A_i\otimes B_i\left| string_j\right\rangle &=&0,\text{ for }i\neq j, 
\nonumber
\end{eqnarray}
where $\left| string_i\right\rangle $ is the state of $i^{\prime }th$
``likely'' string; $\left| string_i^{\prime }\right\rangle $ is the state
after $A_i\otimes B_i$ acts on the $\left| string_i\right\rangle $. If the
state $\left| string_i^{\prime }\right\rangle s$ is an entangled state Alice
and Bob get a yield of entanglement, so the operators $\{A_i\otimes B_i\}$
also work for the distillation of entanglement. On the other hand, if $%
\sigma $ is distillable, as shown in equation (\ref{ed}) there must be
elements $A_i\otimes B_i$ such that as $n\rightarrow \infty ,A_i^{+}\otimes
B_i^{+}\sigma ^{\otimes n}A_i\otimes B_i\rightarrow \left| \Psi
_i\right\rangle \left\langle \Psi _i\right| .$ Each operator $A_i\otimes B_i$
projects out the pure entangled state $\left| \Psi _i\right\rangle .$ If $%
\left| \Psi _i\right\rangle $ belongs to only a string, the operations $%
\{A_i\otimes B_i\}$ for the distillation of entanglement also work for the
local distinguishability of the ``likely'' strings.

To discuss the generalization of the E.q (\ref{dd}) to more general cases,
we should consider two questions: 1. Whether a general measure for the
distillation of entanglement can be carried out by many {\it one by one
measurements }or not{\it ;} 2. It is possible for Alice and Bob to distill a
pure entangled state $\left| \Psi _i\right\rangle $ from the $2^{nS(\rho )}$
strings but not to distinguish each string, so we should revise the E.q (\ref
{dd}). How to revise it? Although the two questions are still open
questions, we believe there are some states the distillable entanglement of
these states can be gained from the E.q (\ref{dd}).

In summary, the transformation of information in the distillation of
entanglement and the locally distinguishability of orthogonal quantum states
plays an important role. In terms of information one can get a general link
between the distillation of entanglement and the distinguishability of
orthogonal quantum states. This link may be useful to calculate distillable
entanglement or get a lower bound of distillable entanglement, and
understand the essence of entanglement \cite{11}.


\begin{references}
\bibitem{1}  C. H. Bennett, D.P. DiVincenzo, C.A. Fuchs, T.Mor, E.Rains,
P.W. Shor, J.A. Smolin, and W.K. Wootters, Phys. Rev. A 59,1070 (1999) or
quant-ph/9804053.

\bibitem{2}  S.Ghosh, G.Kar, A.Roy, A.Sen and U.Sen, Phys.Rev.Lett.87,
277902 (2001)

\bibitem{3}  J.Walgate, A.J.Short, L.Hardy and V.Vedral,
Phys.Rev.Lett.85,4972 (2000); J.Walgate and L.Hardy, Phys.Rev.Lett 89,
127901 (2002); P.-X Chen and C.-Z Li, quant-ph/0209048.

\bibitem{4}  Y.-X.Chen and D.Yang, Phys.Rev.A 64, 064303 (2001)

\bibitem{5}  S. Virmani, M.F. Sacchi, M.B. Plenio and D. Markham, Physics
Letters A 288, 62-68 (2001); M. Horodecki, P. Horodecki, and R. Horodecki,
Acta Physica Slovaca, 48, (1998) 141, or quant-ph/9805072

\bibitem{6}  C. H. Bennett, D. P. Divincenzo, J. A.Smolin, and W.
K.Wootters, Phys. Rev. A{\bf \ 54}, 3824 (1996).

\bibitem{7}  C.H. Bennett and S.J. Wiesner, Phys.Rev.Lett.69,2881 (1992).

\bibitem{8}  C.H. Bennett, G.Brassard, C.Crepeau, R.Jozsa, A.Peres and
W.K.Wootters, Phys.Rev.Lett.70,1895 (1993).

\bibitem{chen}  P.-X.Chen, L.-M Liang, C.-Z Li and M.-Q Huang, Phys.Rev.A65,
012317(2002); Phys.Rev.A66, 022309(2002)

\bibitem{9}  C. H. Bennett, G. Brassard, S. Popescu, B. Schumacher and W. K.
Wootters, Phys. Rev. Lett 76 722 (1996)

\bibitem{66}  David P. Divincenzo, Peter W. Shor and John A. Smolin, Phys.
Rev. A{\bf \ 57}, 830 (1998).

\bibitem{10}  J. Eisert, T. Felbinger, P. Papadopoulos, M.B. Plenio and M.
Wilkens, Phys. Rev. Lett. 84, 1611 (2000); L.Henderson and V.Vedral,
Phys.Rev.Lett 84, 2263 (2000).

\bibitem{12}  G. Vidal and J. I. Cirac, Phys.Rev.Lett 86, 5803 (2001).

\bibitem{v}  V. Vedral and M. B. Plenio, Phys. Rew A 57, 1619 (1998);

\bibitem{m}  V. Vedral M. B. Plenio, K. Jocobs and P. L. Knight, Phys. Rew A
56, 4452 (1997);

\bibitem{b3}  C. H. Bennett, G. Brassard, S. Popescu and B. Schumacher,
Phys. Rev. A{\bf \ 53} 2046 (1996).

\bibitem{b8}  E.M.Rains, Phys.Rew.A 60,173 (1999)

\bibitem{21}  Y.-X.Chen and D.Yang, Phys.Rev.A 66, 014303 (2002)

\bibitem{11}  C.Brukner, M.Zukowski and A.Zeilinger, quant-ph/0106119.
\end{references}
\end{document}